\documentclass[prd,twocolumn,showpacs,amsmath,amssymb,superscriptaddress,floatfix,nofootinbib]{revtex4}
\usepackage{mathrsfs,bm}
\usepackage{longtable,lscape}
\usepackage{txfonts}
\usepackage{amssymb}
\usepackage{indentfirst}
\usepackage{graphicx,,booktabs}
\usepackage{ulem}
\usepackage{multirow}
\usepackage{color}
\usepackage{amssymb}
\definecolor{cover}{rgb}{0.77,0.87,0.88}
\definecolor{blueone}{rgb}{0.1,0.1,.7}
\definecolor{citec}{rgb}{0.14,0.47,0.09}
\definecolor{two}{rgb}{0.0,0.5,0.}
\definecolor{three}{rgb}{.5,.1,0.15}
\usepackage[bookmarks=true,bookmarksopen=false,plainpages=false,breaklinks=true,
   bookmarksnumbered=true,hypertexnames=false,filecolor=blue,urlcolor=blueone,menucolor=three,
   linkcolor=three,citecolor=blueone, colorlinks,anchorcolor=blue,%
   pdfstartview=FitH,pdftitle=title,%
   pdfauthor=author]{hyperref}

\usepackage{relsize}
\usepackage{xspace}
\def\babar{\mbox{\slshape B\kern-0.1em{\smaller A}\kern-0.1em
    B\kern-0.1em{\smaller A\kern-0.2em R}}}

\begin{document}
\title{Radiative decay of $\Xi_b(6227)$ in a hadronic molecule picture}

\author{HongQiang Zhu}
\affiliation{College of Physics and Electronic Engineering, Chongqing Normal University,  Chongqing 401331,China}
\author{Yin Huang \footnote{corresponding author}}
\email{huangy2019@swjtu.edu.cn}
\affiliation{School of Physical Science and Technology, Southwest Jiaotong University, Chengdu 610031,China}

\date{\today}
\begin{abstract}
The $\Xi_b(6227)$ baryon with the quantum number $J^P=1/2^{-}$ is considered as a molecular state composed
of a $\Sigma_b$ baryon and $\bar{K}$ meson.  The partial decay widths of the $\Sigma_b\bar{K}$ molecular state
into $\Xi_b\gamma$ and $\Xi_b^{'}\gamma$ final states through hadronic loops are evaluated with the help of the
effective Lagrangians.  The partial widths for the $\Xi_b(6227)\to\gamma\Xi_b$ and $\Xi_b(6227)\to\gamma\Xi^{'}_b$
are evaluated to be about 1.50-1.02 KeV and 17.56-24.91 KeV, respectively, which may be accessible for the  LHCb.
Based on our results we argue that an experimental determination of the radiative decay width of $\Xi_b(6227)$ is
important for the understanding of its intrinsic properties.
\end{abstract}


\maketitle
\section{INTRODUCTION}

In the past five years, many new narrow bottom baryons were discovered.  The latest one is in 2018, the LHCb
Collaboration observed a new narrow structure with the statistical significance of 9.2 $\sigma$ in the
$\Lambda_b^{0}K^{-}$ and $\Xi_b^{0}\pi^{-}$ invariant mass spectra, named $\Xi_b(6227)^{-}$~\cite{Aaij:2018yqz}.
The observed resonance parameters of the structure are
\begin{align}
&M=6226.9\pm{2.0}(stat)\pm0.3(syst)\pm{}0.2(\Lambda^0_b)~ {\rm MeV}\nonumber\\
&\Gamma=18.1\pm5.4(stat)\pm1.8(syst)~{\rm MeV},
\end{align}
respectively.  The observed channel indicates the isospin of the $\Xi_b(6227)^{-}$ is $1/2$.

However, its spin-parity has not been determined in the experiment.  Assuming different assignments for its
spin-parity, some theoretical interpretations were already discussed in the literature.  Considering $\Xi_b(6227)$
as a conventional bottom baryon, the spin-parity was assigned to be $J^P = 3/2^-$ or $J^P = 5/2^-$ in Refs.~\cite{Chen:2018orb,Wang:2018fjm,Aliev:2018lcs,Chua:2018lfa,Cui:2019dzj}.  Besides supposing $\Xi_b(6227)$ to
be a conventional bottom baryon, the $\Xi_b(6227)$ was explained as a $S-$ wave dynamically generated resonance
with a dominant $\bar{K}\Sigma_b$ configuration~\cite{Yu:2018yxl,Nieves:2019jhp}.   However, a different conclusion
was got that the $\Xi_b(6227)$ can be identified as a pure $\bar{K}\Sigma_b$ bound state with $J^P=1/2^{-}$~\cite{Huang:2018bed}.

To date, the inner structure of this  state remains unclear, and more efforts are necessary.  On the theoretical side,
the study on the decay properties may provide a way of learning about the nature of the $\Xi_b(6227)$.  Regarding the
$\Xi_b(6227)$ as a conventional bottom baryon, the strong decay width has been computed~\cite{Chen:2018orb,Wang:2018fjm,Aliev:2018lcs,Chua:2018lfa,Cui:2019dzj}.  The results indicate that the
$\Xi_b(6227)$ can be well interpreted as a conventional three quark state in comparison with the experimental total
width.   However, the $S-$ wave $\bar{K}\Sigma_b$ assignment for the $\Xi_b(6227)$ is also supported by studying the
strong decay widths~\cite{Yu:2018yxl,Nieves:2019jhp,Huang:2018bed}.  One finds that the two-body allowed strong decay
widths from different model are consistent with each other within errors.   In other word, based on the analysis of
the two-body allowed strong decays, the $\Xi_b(6227)$ is not only can be considered as a conventional three quark
state, but also can be considered as a $S$-wave $\bar{K}\Sigma_b$ molecular.  Theoretical investigations on other
decay modes  will be very helpful to determine whether the $\Xi_b(6227)$ is a conventional bottom baryon or a molecular
state.

The  radiative decays may be helpful to distinguish the internal structure of the $\Xi_b(6227)$, since the coupling
of the photon to the constituent $\bar{K}$ meson and $\Sigma_b$ baryon of the $\Xi_b(6227)$ is essentially different
from that of the quark models for which the photon directly couples to the quark system~\cite{Koniuk:1979vy}.
Therefore, a precise measurements of the radiative decays can be quite useful to test the different interpretations
of the $\Xi_b(6227)$.  However, no work has been done to discuss the radiation decays of the $\Xi(6227)$.  In the
present work we continue our study of the $\Xi_b(6227)$ properties considering its radiative decays in the hadronic
molecule approach developed in our previously paper~\cite{Huang:2018bed}.

This work is organized as follows. The theoretical formalism is explained in Sec. II. The predicted partial
decay widths are presented in Sec. III, followed by a short summary in the last section.

\section{Theoretical FORMALISM}
In our previous paper~\cite{Huang:2018bed}, the $\Xi_b(6227)$ is interpreted as a pure $\bar{K}\Sigma_b$ bound state
with $J^P=1/2^{-}$ by studying the strong decay model, where consistency with the observed strong decay width of the
$\Xi_b(6227)$ was achieved in a hadronic molecule interpretation~\cite{Huang:2018bed}.  According to the pure $\bar{K}\Sigma_b$
molecular scenario,  we calculate the radiative decay widths $\Xi_b(6227)\to\Xi_b\gamma$ and $\Xi_b(6227)\to\Xi_b^{'}\gamma$
in this work, that would help to understand the internal structure of the $\Xi_b(6227)$.  In the $\bar{K}\Sigma_b$ molecular
scenario, the $\Xi_b(6227)$ should couple to its components via $S-$ wave, and the corresponding effective Lagrangian is in
the form~\cite{Huang:2018bed,Dong:2010gu}
\begin{align}
{\cal{L}}_{\Xi_b(6227)}&=g_{\Xi_b(6227)\bar{K}\Sigma_b}\Phi[(k_1\omega_{\Sigma_b}-k_2\omega_{\bar{K}})^2]\nonumber\\
                       &\times\bar{\Xi}_b(6227)\vec{\tau}\cdot\vec{\Sigma}_b\bar{K}\label{eq1},
\end{align}
where $\omega_{ij}=m_i/(m_i+m_{j})$ with $m_i$ is the mass of the baryon or meson.  $k_1$ and $k_2$ is four-momenta
of the $\bar{K}$ meson and $\Sigma_b$ baryon, respectively.  From Refs.~\cite{Huang:2018bed,Dong:2010gu,Faessler:2007gv,Dong:2010xv},
we find  that the correlation function $\Phi(p^2)\doteq\exp(-p_E^2/\Lambda^2)$ is not only introduced to describe
the distributions of the $\bar{K}$ meson and $\Sigma_b$ baryon  in the hadronic molecule but also plays a role to
stop the Feynman diagrams ultraviolet divergence.  The $p_E$ being the Euclidean Jacobi momentum and the $\Lambda$
being the size parameter which characterizes the distribution of the components inside the molecule.  It should be
noted that the $\Lambda$ is a free parameter and it is usually chosen to be about $1$ GeV to reproduce the experimental
observed decay width in the literature~\cite{Huang:2018bed,Dong:2010gu,Faessler:2007gv,Dong:2010xv}.  In this work,
we vary  $\Lambda$ in a range of 0.9 GeV $\leq\Lambda\leq{}1.10$ GeV.
\begin{figure}[h!]
\begin{center}
\includegraphics[bb=100 540 450 720, clip, scale=0.5]{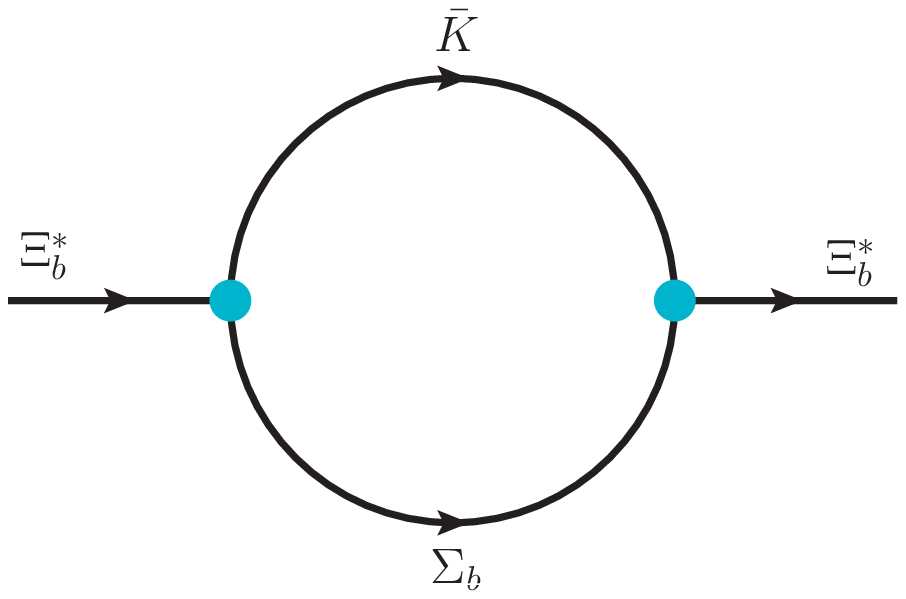}
\end{center}
\caption{(color online). Self-energy of the $\Xi_b^{*}$ state.   }\label{msef}
\end{figure}

In the above Lagrangian,  the remaining unknown coupling constant $g_{\Xi_b(6227)\Sigma_b\bar{K}}$ can be computed
by the compositeness condition~\cite{Weinberg:1962hj,Salam:1962ap}, which indicates that the renormalization constants
of a composite particle wave function should be zero, i.e.,
\begin{align}
Z_{\Xi_b(6227)}=1-\frac{d{\cal{S}}[\Xi_b(6227)]}{dk\!\!\!/_0}|_{k\!\!\!/_0=m_{\Xi_{b}(6227)}}=0\label{eqn3},
\end{align}
where the ${\cal{S}}[\Xi_b(6227)]$ is the mass operator of the $\Xi_b(6227)$ corresponding to the diagrams in Fig.~\ref{msef}.
With the effective Lagrangian in Eq.~(\ref{eq1}) and the compositeness condition in Eq.~(\ref{eqn3}), we can obtain the coupling
constant of the $\Sigma_b\bar{K}$ molecule to its components
\begin{align}
\frac{1}{g^2_{\Xi_b(6227)\Sigma_b\bar{K}}}&=\sum_{i=1}^{2}{\cal{A}}^2_{i}\int_{0}^{\infty}d\alpha\int_{0}^{\infty}d\beta\frac{1}{16\pi^2z^2}\nonumber\\
                                          &\times[-\frac{\Delta_i}{2z}-\frac{2}{\Lambda^2}m_{\Xi_b^{*}}({\cal{M}}_{i}-\frac{\Delta_im_{\Xi_b^{*}}}{2z}){\cal{F}}_i]\nonumber\\
                                          &\times\exp\{-\frac{1}{\Lambda^2}[{\cal{F}}_im^2_{\Xi_b^{*}}+\alpha{}{\cal{M}}_i^2+\beta{}m_i^2]\}\label{coup},
\end{align}
where ${\cal{F}}_i=-2\omega_i^2+\frac{\Delta^2_i}{4z}-\beta$ and $\Delta_i=-4\omega_i-2\beta$ with $i=1$ and $2$
denoting the molecule component $K^{-}\Sigma_b^0$ and $\bar{K}^{0}\Sigma_b^{-}$, respectively.  ${\cal{M}}_i$ and $m_i$
are the mass of the $\Sigma_b$ baryon and the mass of the $\bar{K}$ meson, respectively.  The isospin-spin symmetry
requires the coupling of the $\Xi_b^{-}\Sigma_b^{-}\bar{K}^{0}$ vertex is $\sqrt{2}$ times lager than the one of the
$\Xi_b^{-}\Sigma_b^{0}K^{-}$.  Simply employing those ratios and the ${\cal{A}}_i$ listed
\begin{align}
{\cal{A}}_i=\left\{
\begin{aligned}
\sqrt{2} &  & {i=\Sigma_b^{-}\bar{K}^{0}} \\
1        &  & {i=\Sigma_b^{0}K^{-}}.
\end{aligned}
\right.
\end{align}

\begin{figure}[h!]
\begin{center}
\includegraphics[bb=80 400 600 710, clip, scale=0.40]{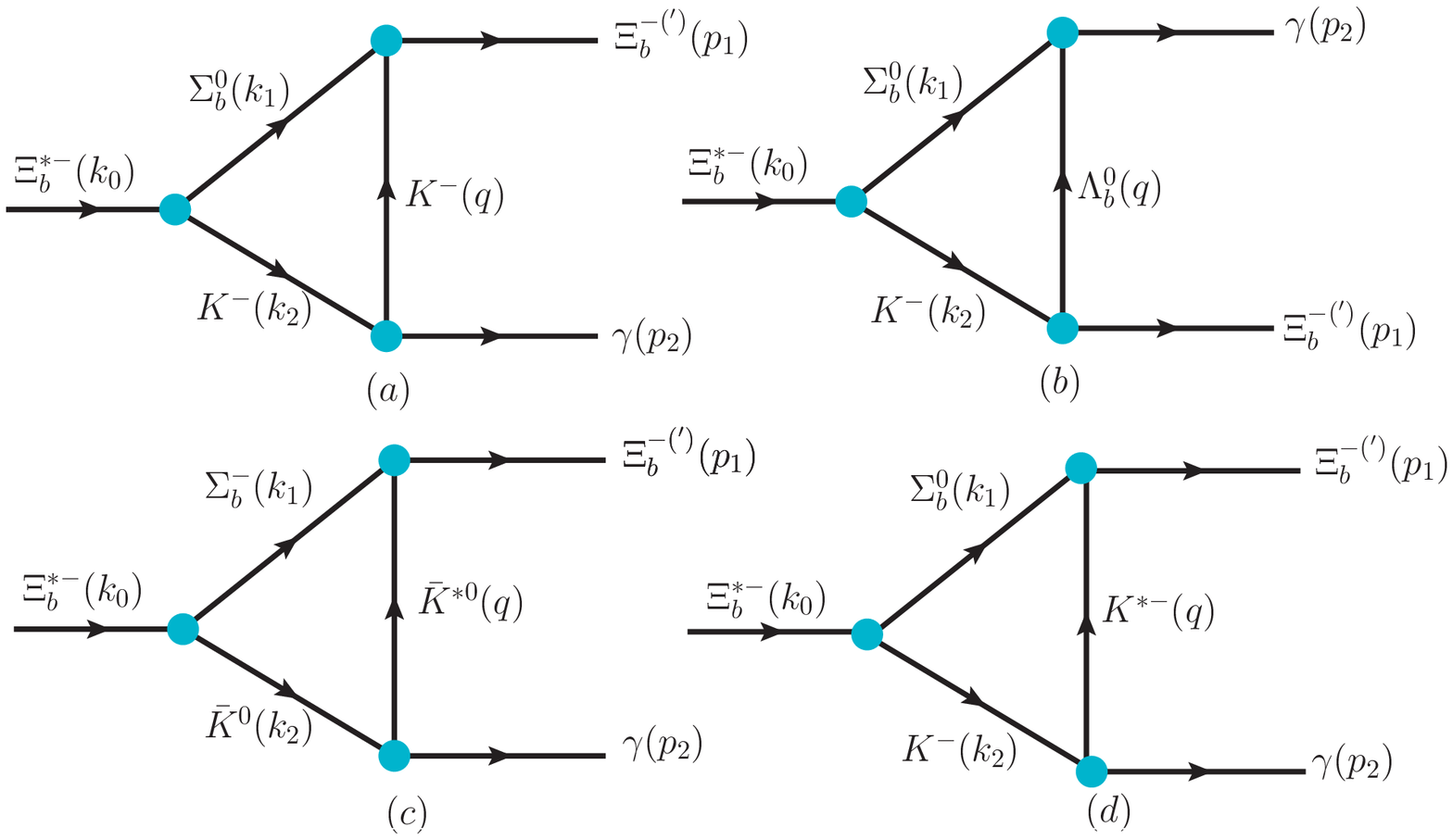}
\end{center}
\caption{(color online) Feynman diagrams for the $\Xi_b(6227)^{-}\to\Xi_b^{-}\gamma$ and $\Xi_b(6227)^{-}\to\Xi_b^{'-}\gamma$
decay processes.   We also show the definitions of the kinematics ($k_0,k_1,k_2,p_1,p_2$, and $q$) used in the calculation.   }\label{fety}
\end{figure}

In the present hadronic molecular scenario, the diagrams contributing to the $\Xi_b(6227)^{-}\to\Xi^{-}_b\gamma$ and
$\Xi_b(6227)^{-}\to\Xi_b^{'-}\gamma$ decay are presented in
Fig.~\ref{fety}.

In order to compute the radiative decays of the diagrams shown in Fig.~\ref{fety}, the effective Lagrangian densities
related to the photon fields are needed, which are~\cite{Chen:2010re,Kim:2014hha}
\begin{align}
{\cal{L}}_{K^{*}K\gamma}&=\frac{g_{K^{*+}K^{+}\gamma}}{4}e\epsilon^{\mu\nu\alpha\beta}F_{\mu\nu}K^{*+}_{\alpha\beta}K^{-}\nonumber\\
                        &+\frac{g_{K^{*0}K^{0}\gamma}}{4}e\epsilon^{\mu\nu\alpha\beta}F_{\mu\nu}K^{*0}_{\alpha\beta}\bar{K}^{0}+h.c.,\label{eqw1}\\
{\cal{L}}_{KK\gamma}&=ieA_{\mu}K^{-} \overleftrightarrow{\partial}^{\mu}K^{+}\label{eqw2},\\
{\cal{L}}_{\gamma{}\Sigma_b\Lambda_b^0}&=\frac{e\mu_{\Sigma_b\Lambda_b}}{2m_{\Lambda_b^0}}\bar{\Sigma}_b^0\sigma_{\mu\nu}\partial^{\nu}A^{\mu}\Lambda_b^0+H.c.\label{eqw3},
\end{align}
where the strength tensor are defined as $F_{\mu\nu}=\partial_{\mu}A_{\nu}-\partial_{\nu}A_{\mu}$ and $K^{*}_{\mu\nu}=\partial_{\mu}K^{*}_{\nu}-\partial_{\nu}K^{*}_{\mu}$.
The $\alpha=e^2/4\pi=1/137$ is the electromagnetic fine structure constant.  The $\mu_{\Sigma_b\Lambda_b}=-1.37{}\mu$
is transition magnetic moment~\cite{Wang:2018cre},  which $\mu$ is in the unit of the nuclear magneton.  The coupling
constant $g_{K^{*+}K^{+}\gamma}$ and $g_{K^{*0}K^{0}\gamma}$ can be determined from the partial decay width of
$K^{*+}\to{}K^{+}\gamma$ and $K^{*0}\to{}K^{0}\gamma$, which can be obtained from Eq.~(\ref{eqw1}),
\begin{align}
&\Gamma(K^{*+}\to{}K^{+}\gamma)=\frac{\alpha{}g^2_{K^{*+}K^{+}\gamma}}{24}m_{K^{*+}}(m^2_{K^{*+}}-m^2_{K^{+}}),\\
&\Gamma(K^{*0}\to{}K^{0}\gamma)=\frac{\alpha{}g^2_{K^{*0}K^{0}\gamma}}{24}m_{K^{*0}}(m^2_{K^{*0}}-m^2_{K^{0}})\label{eq10},
\end{align}
where $m_{K^{*}}$ and $m_{K}$ are mass of $K^{*}$ and kaon, respectively.

According to the experimental widths $\Gamma(K^{*+}\to{}K^{+}\gamma)=0.0503$ KeV~\cite{Tanabashi:2018oca}, $\Gamma(K^{*0}\to{}K^{0}\gamma)=0.125$ KeV~\cite{Tanabashi:2018oca},
and the masses of the particles that shown in Table~\ref{table1}, the coupling constant $g_{K^{*}K\gamma}$ is fixed as
\begin{align}
g_{K^{*+}K^{+}\gamma}=0.580~ {\rm GeV^{-1}},~~~~~g_{K^{*0}K^{0}\gamma}=-0.904~ {\rm GeV^{-1}}.
\end{align}
The signs of these coupling constants are fixed by the quark model.

To evaluate the diagrams in Fig.~\ref{fety}, in addition to the Lagrangian in Eqs.(~\ref{eq1},~\ref{eqw1},~\ref{eqw2}),
the following effective Lagrangians, responsible for  heavy baryons and pseudoscalar mesons interactions are needed
as well~\cite{Meng:2018gan}
\begin{align}
{\cal{L}}_{B\phi}&=g_1\langle\bar{B}_6\gamma_{\mu}\gamma_5u^{\mu}B_{6}\rangle+g_2\langle\bar{B}_6\gamma_{\mu}\gamma_5u^{\mu}B_{\bar{3}}+H.c.\rangle\nonumber\\
                 &+g_6\langle\bar{B}_{\bar{3}}\gamma_{\mu}\gamma_5u^{\mu}B_{\bar{3}}\rangle \label{qer1}.
\end{align}
Where the coupling constant $g_1=-\sqrt{\frac{8}{3}}g_2$ and $g_6=0$~\cite{Meng:2018gan}.  $u^{\mu}$ is the axial
vector combination of the pseudoscalar-meson fields and its derivatives,
\begin{align}
u^{\mu}=i(u^{\dagger}\partial^{\mu}u-u\partial^{\mu}u^{\dagger}),
\end{align}
where the $u^2=U=\exp(i\frac{\phi}{f_0})$, $f_0$=92.4 MeV, and the pseudoscalar-meson octet $\phi$ are represented
by the $3\times3$ matrix
\begin{equation}
\phi=\sqrt{2}
\left(
  \begin{array}{ccc}
    \frac{\pi^{0}}{\sqrt{2}}+\frac{\eta}{\sqrt{6}} &    \pi^{+}                                        &       K^{+}\\
    \pi^{-}                                       &    -\frac{\pi^{0}}{\sqrt{2}}+\frac{\eta}{\sqrt{6}} &       K^{0}\\
    K^{-}                                         &    \bar{K}^{0}                                     &       -\frac{2}{\sqrt{6}}\eta
  \end{array}
\right)\label{eq7}.
\end{equation}
The particle assignment for the $J = 1/2$ bottom baryons of the $B_{\bar{3}}$ and $B_{6}$ representations
is~\cite{Meng:2018gan,Banuls:1999br}
\begin{equation}
B_{\bar{3}}=
\left(
  \begin{array}{ccc}
    0                                             &~~~~~    \Lambda_b^0                                     &~~~~~       \Xi_b^{0}\\
    \\
    -\Lambda_b^0                                  &~~~~~    0                                               &~~~~~       \Xi_b^{-}\\
    \\
    -\Xi_b^{0}                                    &~~~~~    -\Xi_b^{-}                                      &~~~~~       0
  \end{array}
\right)\label{eq8},
\end{equation}
\begin{equation}
B_{6}=
\left(
  \begin{array}{ccc}
    \Sigma_b^{+}                      &~~~~~    \sqrt{\frac{1}{2}}\Sigma_b^0      &~~~~~        \sqrt{\frac{1}{2}}\Xi_b^{0'} \\
    \sqrt{\frac{1}{2}}\Sigma_b^0      &~~~~~    \Sigma_b^{-}                      &~~~~~        \sqrt{\frac{1}{2}}\Xi_b^{-'}\\
    \sqrt{\frac{1}{2}}\Xi_b^{0'}      &~~~~~    \sqrt{\frac{1}{2}}\Xi_b^{-'}      &~~~~~         \Omega_b^{-}
  \end{array}
\right)\label{eq9}.
\end{equation}
\begin{table}[h!]
\centering
\caption{Masses of the particles needed in the present work (in units of MeV).}\label{table1}
\begin{tabular}{cccccccccc}
\hline\hline
$\Sigma_b^{+}$ &$\Sigma_b^{0}$  &$\Sigma_b^{-}$  &$\Lambda^{0}_b$   &$\Xi_b^{-}$   & $\Xi_b^{'-}$          \\
$5811.3$       &$5813.4$        &$5815.5$        &$5619.6$          &$5794.5$      & $5935.02   $        \\ \hline
$K^{0}$        &$K^{*0}$        &$K^{*\pm}$      &$K^{\pm}$         &$\pi^{\pm}$   &                      \\
$497.611$      &$898.36$        &$891.66$        &$493.68$          &$139.57$      &                     \\ \hline \hline
\end{tabular}
\end{table}

The coupling $g_2$ is fixed from the strong decay width of $\Sigma_b\to{}\Lambda_b^0\pi$.  With the help of
Eqs.~(\ref{qer1}-\ref{eq9}), the two-body decay width $\Gamma(\Sigma_b\to{}\Lambda_b^0\pi)$ is related to $g_2$ as
 \begin{align}
 \Gamma(\Sigma_b\to{}\Lambda_b^0\pi)=\frac{g_2^2}{\pi{}f_0^2}\frac{(m_{\Sigma_b}+m_{\Lambda_b^0})^2}{(m_{\Sigma_b}+m_{\Lambda_b^0})^2-m^2_{\pi}}{\cal{P}}^3_{\pi{}\Lambda_b^{0}},
 \end{align}
where the $m_{\Sigma_b}$, $m_{\Lambda_b^0}$, and $m_{\pi}$ are the masses of the $\Sigma_b$ baryon, $\Lambda_b^{0}$
baryon, and $\pi$ meson, respectively.   The ${\cal{P}}_{\pi{}\Lambda_b^0}$ is the three-momentum of the $\pi$ in the
rest frame of the $\Sigma_b$.   Particle Data Group list the $\Sigma_b$ predominantly decays into $\Lambda_b^0\pi$~\cite{Tanabashi:2018oca},
so its partial width may be approximately equal to the total width of the $\Sigma_b$.  Using the experimental strong
decay width and the masses of the particles needed in the present work we obtain $g_2=0.252\pm{}0.013$.

To compute the radiative decay amplitude, the $K^{*-}\Sigma^{0}_b\Xi_b^{-(')}$ and $\bar{K}^{*0}\Sigma^{-}_b\Xi_b^{-(')}$
vertices Lagrangian are also needed.  Following the strategy of Ref.~\cite{Hofmann:2005sw,Montana:2017kjw}, the Lagrangian
for the $K^{*-}\Sigma^{0}_b\Xi_b^{-(')}$ and $\bar{K}^{*0}\Sigma^{-}_b\Xi_b^{-(')}$ vertexes  can be easily obtained by
replacing the charm baryons with the bottom ones, whose procedures are just illustrated in Ref.~\cite{Huang:2018bed},
\begin{align}
{\cal{L}}_{K^{*}\Sigma_b\Xi_b^{(')}}&=\frac{g}{\sqrt{6}}\bar{\Xi}^{-}_b\gamma^{\mu}\bar{K}_{\mu}^{*0}\Sigma_b^{-}+\frac{g}{2\sqrt{3}}\bar{\Xi}^{-}_b\gamma^{\mu}K_{\mu}^{*-}\Sigma_b^{0}\nonumber\\
                                    &+\frac{g}{\sqrt{6}}\bar{\Xi}^{0}_b\gamma^{\mu}K^{*-}_{\mu}\Sigma_b^{+}-\frac{g}{2\sqrt{3}}\bar{\Xi}^{0}_b\gamma^{\mu}\bar{K}_{\mu}^{*0}\Sigma_b^{0}\nonumber\\
                                    &-\frac{g}{\sqrt{2}}\bar{\Xi}^{'0}_b\gamma^{\mu}K^{*-}_{\mu}\Sigma_b^{+}+\frac{g}{2}\bar{\Xi}^{'0}_b\gamma^{\mu}\bar{K}_{\mu}^{*0}\Sigma_b^{0}\nonumber\\
                                    &+\frac{g}{\sqrt{2}}\bar{\Xi}^{'-}_b\gamma^{\mu}\bar{K}_{\mu}^{*0}\Sigma_b^{-}+\frac{g}{2}\bar{\Xi}^{'-}_b\gamma^{\mu}K_{\mu}^{*-}\Sigma_b^{0}+H.c.,                               \end{align}
where the coupling constant $g=6.6$ and is get from Ref.~\cite{Montana:2017kjw}.

In the radiative decay of the $\Xi_b(6227)^{-}\to\Xi_b^{-}\gamma$ and $\Xi_b(6227)^{-}\to\Xi_b^{'-}\gamma$, the photon can be
emitted from the kaon meson and $\Sigma_b$ baryon.  The triangle diagrams are listed in Fig.~\ref{fety} and the corresponding
amplitudes are
\begin{align}
{\cal{M}}_{a}&(\Xi_b(6227)^{-}\to\Xi_b^{-}\gamma)=-(i)^3\frac{eg_{\Xi_b^{*}\Sigma_b\bar{K}}g_2}{f_0}\int\frac{d^4q}{(2\pi)^4}\nonumber\\
                          &\times\Phi[(k_1\omega_{K^{-}}-k_2\omega_{\Sigma_b^{0}})^2]\bar{u}(p_1)q\!\!\!/\gamma_5\frac{k\!\!\!/_1+m_{\Sigma_b^{0}}}{k_1^2-m^2_{\Sigma^{0}_b}}u(k_0)\nonumber\\
                          &\times\frac{1}{k_2^2-m^2_{K^{-}}}(k_2^{\mu}+q^{\mu})\frac{1}{q^2-m^2_{K^{-}}}\epsilon^{*}_{\mu}(p_2),\nonumber\\
{\cal{M}}_{b}&(\Xi_b(6227)^{-}\to\Xi_b^{-}\gamma)=0,\nonumber\\
{\cal{M}}_{c}&(\Xi_b(6227)^{-}\to\Xi_b^{-}\gamma)=-(i)^3\frac{geg_{\Xi_b^{*}\Sigma_b\bar{K}}g_{\bar{K}^{*0}\bar{K}^{0}\gamma}}{4\sqrt{3}}\nonumber\\
                          &\times\int\frac{d^4q}{(2\pi)^4}\Phi[(k_1\omega_{\bar{K}^0}-k_2\omega_{\Sigma_b^{-}})^2]\bar{u}(p_1)\gamma_{\eta}\frac{k\!\!\!/_1+m_{\Sigma_b^{-}}}{k_1^2-m^2_{\Sigma^{-}_b}}\nonumber\\
                          &\times{}u(k_0)\frac{1}{k_2^2-m^2_{\bar{K}^{0}}}\epsilon_{\rho\nu\alpha\beta}(p_2^{\rho}g^{\nu\mu}-p_{2}^{\nu}g^{\rho\mu})\nonumber\\
                          &\times(q^{\alpha}g^{\beta\sigma}-q^{\beta}g^{\alpha\sigma})(-g^{\eta\sigma}+\frac{q^{\eta}q^{\sigma}}{m^2_{\bar{K}^{*0}}})\frac{1}{q^2-m^2_{\bar{K}^{*0}}}\epsilon^{*}_{\mu}(p_2),\nonumber\\
{\cal{M}}_{d}&(\Xi_b(6227)^{-}\to\Xi_b^{-}\gamma)=-(i)^3\frac{geg_{\Xi_b^{*}\Sigma_b\bar{K}}g_{K^{*-}K^{-}\gamma}}{8\sqrt{3}}\nonumber\\
                          &\times\int\frac{d^4q}{(2\pi)^4}\Phi[(k_1\omega_{K^{-}}-k_2\omega_{\Sigma_b^{0}})^2]\bar{u}(p_1)\gamma_{\eta}\frac{k\!\!\!/_1+m_{\Sigma_b^{0}}}{k_1^2-m^2_{\Sigma^{0}_b}}\nonumber\\
                          &\times{}u(k_0)\frac{1}{k_2^2-m^2_{K^{-}}}\epsilon_{\eta\nu\alpha\beta}(p_2^{\rho}g^{\nu\mu}-p_{2}^{\nu}g^{\mu\rho})\nonumber\\
                          &\times(q^{\alpha}g^{\beta\sigma}-q^{\beta}g^{\alpha\sigma})(-g^{\eta\sigma}+\frac{q^{\eta}q^{\sigma}}{m^2_{K^{*-}}})\frac{1}{q^2-m^2_{K^{*-}}}\epsilon^{*}_{\mu}(p_2),\nonumber\\
{\cal{M}}_{a}&(\Xi_b(6227)^{-}\to\Xi_b^{'-}\gamma)=\frac{1}{\sqrt{2}}{\cal{M}}_{a}(\Xi_b^{*-}\to\Xi_b^{-}\gamma)|^{g_2\to-g_1}_{m_{\Xi^{-}_b}\to{}m_{\Xi_b^{'-}}},\nonumber
\end{align}

\begin{align}
{\cal{M}}_{b}&(\Xi_b(6227)^{-}\to\Xi_b^{'-}\gamma)=(i)^3\frac{eg_2g_{\Xi_b^{*}\Sigma_b\bar{K}}\mu_{\Sigma_b\Lambda_b}}{4f_0m_{\Lambda_b^0}}\int\frac{d^4q}{(2\pi)^4}\nonumber\\
                          &\times\Phi[(k_1\omega_{K^{-}}-k_2\omega_{\Sigma_b^{0}})^2]\bar{u}(p_1)k\!\!\!/_2\gamma_5\frac{q\!\!\!/+m_{\Lambda_b^0}}{q^2-m^2_{\Lambda_b^0}}\nonumber\\
                          &\times(\gamma_{\mu}p\!\!\!/_2-p\!\!\!/_2\gamma_{\mu})\frac{k\!\!\!/_1+m_{\Sigma_b^0}}{k_1^2-m^2_{\Sigma_b^0}}u(k_0)\frac{1}{k_2^2-m^2_{K^{-}}}\epsilon^{*}_{\mu}(p_2),\nonumber\\
{\cal{M}}_{c}&(\Xi_b(6227)^{-}\to\Xi_b^{'-}\gamma)\nonumber\\
                   &=\sqrt{3}{\cal{M}}_{c}(\Xi_b(6227)^{-}\to\Xi_b^{-}\gamma)|_{m_{\Xi_b^{-}}\to{}m_{\Xi_b^{'-}}},\nonumber\\
{\cal{M}}_{d}&(\Xi_b(6227)^{-}\to\Xi_b^{'-}\gamma)\nonumber\\
                   &=\sqrt{3}{\cal{M}}_{d}(\Xi_b(6227)^{-}\to\Xi_b^{-}\gamma)|_{m_{\Xi_b^{-}}\to{}m_{\Xi_b^{'-}}}.\label{eq19}
\end{align}
The total amplitude of $\Xi_b(6227)^{-}\to\Xi_b^{-(')}\gamma$ is
\begin{align}
{\cal{M}}^{T}_{\Xi_b(6227)^{-}\to\Xi_b^{-(')}\gamma}={\cal{M}}_a+{\cal{M}}_b+{\cal{M}}_c+{\cal{M}}_d.
\end{align}
After performing the loop integral, the total contributions of the triangle diagrams to the
$\Xi_b(6227)^{-}\to\Xi_b^{-(')}\gamma$  can be parameterized as
\begin{align}
{\cal{M}}^{T}_{\Xi_b(6227)^{-}\to\Xi_b^{-(')}\gamma}&=\epsilon^{*}_{\mu}(p_2)\bar{u}(p_1)\nonumber\\
                                                    &\times(g_1^{Tri}\gamma^{\mu}+g_2^{Tri}\frac{p_1^{\mu}p\!\!\!/_{2}}{p_1\cdot{}p_2})u(k_0).
\end{align}
\begin{figure}[h!]
\begin{center}
\includegraphics[bb=100 560 450 720, clip, scale=0.45]{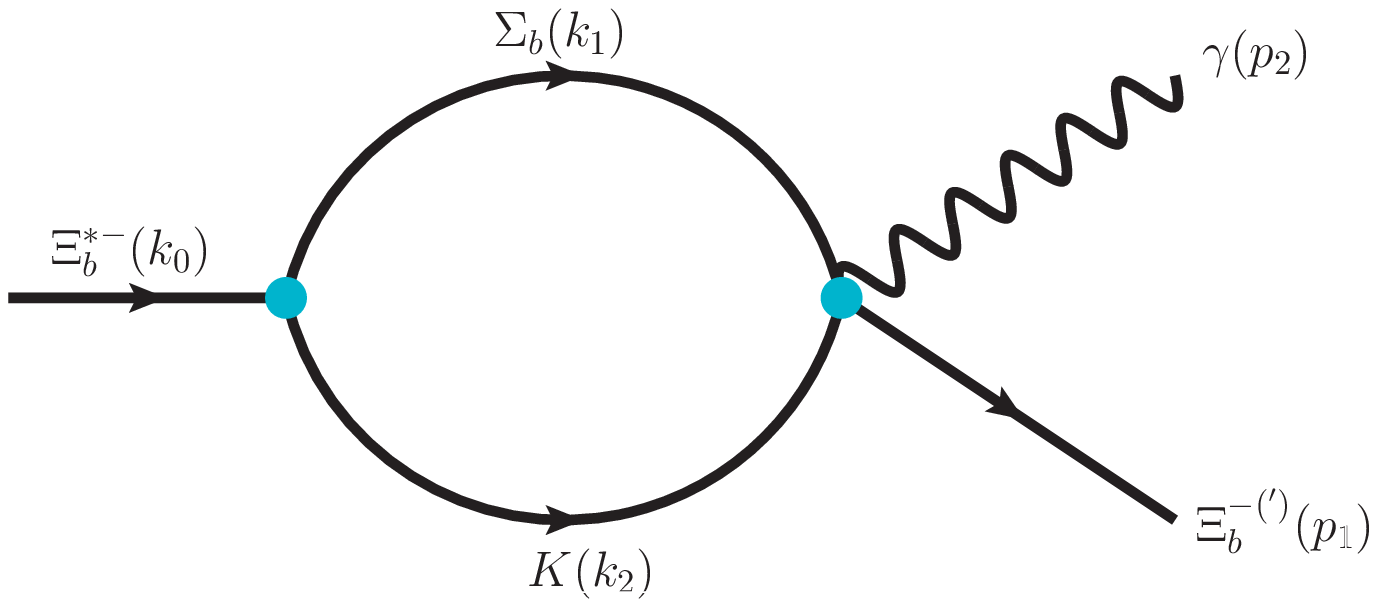}
\end{center}
\caption{(color online) The contact diagram for $\Xi_b(6227)^{-}\to\Xi_b^{-}\gamma$ and $\Xi_b(6227)^{-}\to\Xi_b^{'-}\gamma$.
  We also show the definitions of the kinematics ($k_0,k_1,k_2$, and $q$) used in the calculation.   }\label{fetc}
\end{figure}

One should notice that this amplitude can not satisfies the gauge invariance of the photon field if only triangle diagrams
are included.  To ensure the gauge invariance of the total amplitudes,  the contact diagram in Fig.~\ref{fetc} should be
included.   The effective Lagrangian describing vertex of $\Xi_b^{(')}\gamma\Sigma_b\bar{K}$ could be deduced from the one
of the $\Xi_b^{(')}\Sigma_b\bar{K}$ by minimal substitution $\partial^{\mu}\to\partial^{\mu}+ie{\cal{A}}^{\mu}$, which is
\begin{align}
{\cal{L}}_{\Xi_b^{(')}\gamma\Sigma_b\bar{K}}=g_{\Xi_b^{(')}\gamma\Sigma_b\bar{K}}\bar{\Xi}_b^{(')}\gamma_{\mu}\gamma_5{\cal{A}}^{\mu}\bar{K}\Sigma_b.
\end{align}
By this effective Lagrangian, the amplitude of the contact diagram is in the form,
\begin{align}
{\cal{M}}_{Con}&[\Xi_b(6227)^{-}\to\Xi_b^{-(')}\gamma]=g_{\Xi_b^{(')}\gamma\Sigma_b\bar{K}}\int\frac{d^4k_1}{(2\pi)^4}\nonumber\\
&\times\epsilon^{*}_{\mu}(p_2)\Phi[(k_1\omega_{K^{-}}-k_2\omega_{\Sigma_b^{0}})^2]\bar{u}(p_1)\gamma_{\mu}\gamma_5\nonumber\\
&\times{}\frac{k\!\!\!/_1+m_{\Sigma_b^{0}}}{k_1^2-m^2_{\Sigma^{0}_b}}u(k_0)\frac{1}{k_2^2-m^2_{K^{-}}}.
\end{align}
After performing the loop integral in the above amplitude, we get,
\begin{align}
{\cal{M}}_{Con}&[\Xi_b(6227)^{-}\to\Xi_b^{-(')}\gamma]\nonumber\\
                &=\epsilon^{*}_{\mu}(p_2)\bar{u}(p_1)g^{}_{Con}\gamma^{\mu}u(k_0).
\end{align}

The total amplitude of the $\Xi_b(6227)^{-}\to\Xi_b^{-(')}\gamma$ is the sum of the triangle diagrams
contributions and the contact term contributions.  In order to keep the full amplitude gauge invariant,
the $g^{}_{Con}=-g_1^{Tri}-g_2^{Tri}$ can be got and the coupling constant $g_1^{Tri}$ and $g_2^{Tri}$
could be evaluated from the amplitudes listed in Eqs.~\ref{eq19}.

Once the amplitudes are determined, the corresponding partial decay widths can be obtained, which read,
\begin{align}
\Gamma(\Xi_b(6227)\to MB)=\frac{1}{2J+1}\frac{1}{8\pi}\frac{|\vec{p}_1|}{m^2_{\Xi_b(6227)}}\overline{|{\cal{M}}|^2},
\end{align}
where $J$ is the total angular momentum of the $\Xi_b(6227)$ state, the $|\vec{p}_1|$ is the three-momenta
of the decay products in the center of mass frame, the overline indicates the sum over the polarization vectors
of the final hadrons, and $MB$ denotes the decay channel of $MB$, i.e.,$\Xi_b\gamma$ and $\Xi_b^{'}\gamma$.

\section{RESULTS AND DISCUSSIONS}
\begin{figure}[h!]
\begin{center}
\includegraphics[bb=60 5 800 520, clip,scale=0.33]{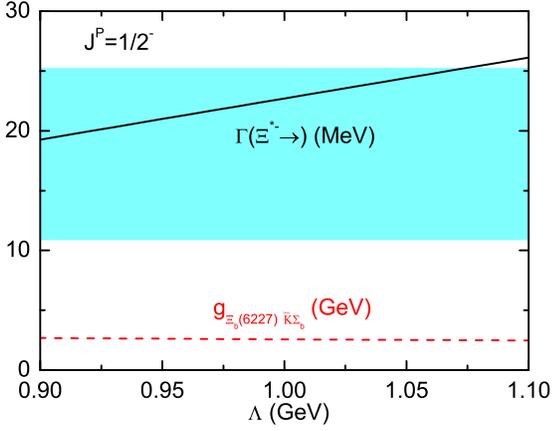}
\caption{(color online). Coupling constant(red dash line),$g_{\Xi_b(6227)\Sigma_b\bar{K}}$, and total decay width(black solid line)
as a function of the parameter $\Lambda$. The oycn bands denote the experimental total width~\cite{Aaij:2018yqz}.} \label{coup-th}
\end{center}
\end{figure}

To estimate the radiative decay widths of the considered processes, the relevant coupling constants $g_{\Xi_b(6227)\Sigma_b\bar{K}}$
should be first discussed.   Regarding the $\Xi_b(6227)$ as $S-$wave loosely $\Sigma_b\bar{K}$ hadronic molecule,  the coupling constant
$g_{\Xi_b(6227)\Sigma_b\bar{K}}$ can be computed via compositeness condition.   As shown in Eq.~\ref{coup}, the coupling constant is
dependent on the parameter $\Lambda$.  With a value of the cutoff $\Lambda=0.9-1.1$ GeV, the corresponding coupling constant is varying
from 2.68 GeV to 2.47 GeV, which is shown in Fig.~\ref{coup-th}.   We note that the coupling constant decreases slowly with the increase
of the cut-off, and  the coupling constant is almost independent of $\Lambda$.   In our previous paper~\cite{Huang:2018bed}, the coupling
constant is incorrect and is smaller than that given in this paper.  However, it does not change the conclusion that $\Xi_b(6227)$ only
can  be  considered as $S-$wave $\bar{K}\Sigma_b$ molecule and the new result is shown in Fig.~\ref{coup-th}.

\begin{figure}[h!]
\begin{center}
\includegraphics[bb=-20 50 800 650, clip,scale=0.40]{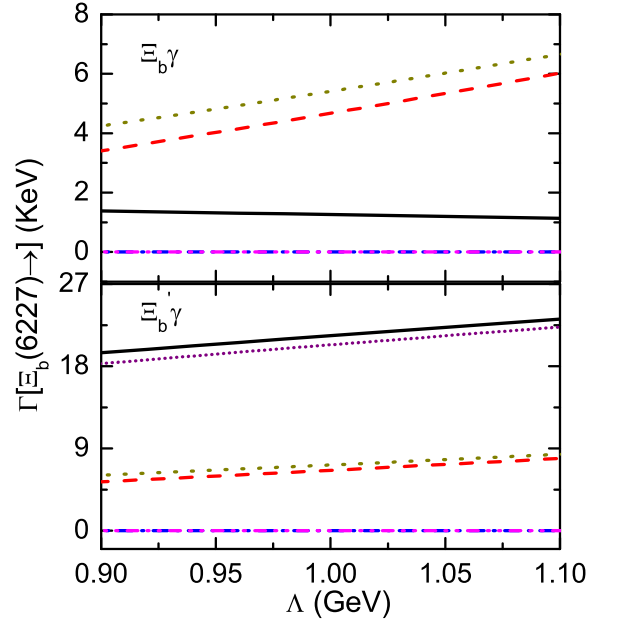}
\caption{(color online). Total decay width (black solid line) and  partial decay widths from $K^{-}$(red dash line),
$\bar{K}^{*0}$(blue dash dot line), $\bar{K}^{*-}$(magenta dash dot dot line), $\Lambda_b^{0}$(purple short line), and
rest is contact term exchange contribution for the $\Xi_b(6227)$ as a function of the parameter $\Lambda$.} \label{width-ty}
\end{center}
\end{figure}

In Fig.~\ref{width-ty}, the dependence of the radiative decay widths on the cutoff $\Lambda$ is shown. With the increasing of
cut-off from 0.9 GeV to 1.1 GeV, the radiative decay width of the $\Xi_b(6227)\to\gamma\Xi^{'}_b$ monotonously increases,  but
very slowly.   In particular, the partial width varies from 19.47 to 23.15 KeV with the variation of $\Lambda$ from 0.9 to 1.1 GeV.
However, the radiative decay widths decreases for the $\Xi_b(6227)\to\gamma\Xi_b$ when we change the cut-off $\Lambda$ from 0.9
to 1.1 GeV.  With the constrained parameter $\Lambda$, the partial width of the $\Xi_b(6227)\to\gamma\Xi_b$ is estimated to be
\begin{align}
\Gamma(\Xi_b(6227)\to\gamma\Xi_b)=1.38-1.14~ {\rm KeV},
\end{align}
which is very weakly dependent on the model parameter.   Our calculation indicates that the width of $\Xi_b(6227)\to\gamma\Xi^{'}_b$
is about one order larger than the one of $\Xi_b(6227)\to\gamma\Xi_b$.

The contribution to the total width coming from individual channels are also shown in Fig.~\ref{width-ty}.  Since the
relative signs of the corresponding amplitude in Fig.~\ref{fety} are well defined, this makes us easy to analyze the
character of the interferences between the various channels.  The total decay widths obtained are the square of their
coherent sum.  As for the $\Xi_b(6227)\to\gamma\Xi_b$ process, the $K^{-}$-exchange and contract term provide
a dominant contributions to the total decay width, and the amplitude of these two channels are not gauge invariant.
The amplitudes corresponding to $\Lambda_b^0$-exchange and $\bar{K}^{*}$-exchange are gauge invariant and the number
results indicate that the contribution of these amplitudes are almost three order smaller than that of the amplitudes
corresponding to $K^{-}$-exchange and contract term.   However, the interferences among them are sizable, which make
the total radiative decay width smaller than the partial decay widths of the $K^{-}$-exchange and contact term, respectively.
For the transition $\Xi_b(6227)\to\gamma\Xi^{'}_b$, the individual contributions of the $\Lambda_b^0$ exchange provides a
dominant over the others, which almost equal to the total width.   One finds that the estimated these individual partial
decay widths are all insensitive to the $\Lambda$.  It should be noted that the radiative decay width of the $\Xi_b(6227)$
into $\gamma\Xi_b$ final state is about one order smaller than into $\gamma\Xi^{'}_b$ final state due to the contribution
of Fig.~\ref{fety}(b) vanishes  in the case of $\Xi_b(6227)\to\gamma\Xi_b$.

According to the Eq.~\ref{eq10} and the experimental width $\Gamma(K^{*0}\to{}K^0\gamma)$=0.125 KeV, the coupling constant
$g_{K^{*0}K^{0}\gamma}$=0.904 GeV$^{-1}$ is fixed.   However, it is usual chosen as about -0.904 GeV$^{-1}$ in the literature
and the sign of this coupling constant is fixed by the quark model.  If we take $g_{K^{*0}K^{0}\gamma}$=0.904 GeV$^{-1}$,
the decay width $\Gamma_{\gamma{}\Xi_b}$=1.39-1.15 KeV and  $\Gamma_{\gamma{}\Xi^{'}_b}$=19.33-23.08 KeV are very close to
previous results:$\Gamma_{\gamma{}\Xi_b}$=1.38-1.14 KeV and  $\Gamma_{\gamma{}\Xi^{'}_b}$=19.47-23.15 KeV.  In other words,
our numerical results show that the value of the decay width $\Gamma_{\gamma{}\Xi_b}$ and $\Gamma_{\gamma{}\Xi^{'}_b}$ are
not very sensitive to the negative sign  when varying the model parameter $\Lambda$ from 0.9 to 1.1 .

\section{summary}
At present, there in no sufficient experimental information to determine the spin-parity of $\Xi_b(6227)$ state.
The study of its decay behaviors at some experimental could provide a further test to the inner structure.  However,
the strong decay channels are hard to determine whether the $\Xi_b(6227)$ is a conventional bottom baryon~\cite{Chen:2018orb,Wang:2018fjm,Aliev:2018lcs,Chua:2018lfa,Cui:2019dzj} or a molecular state~\cite{Yu:2018yxl,Nieves:2019jhp,Huang:2018bed}.

In the present work, we estimated the partial widths for the radiative from the $\Xi_b(6227)$ to the $\Xi^{'}_b$ and $\Xi_b$
in a molecular scenario, in which the $\Xi_b(6227)$ is assigned as a $\Sigma_b{}\bar{K}$ hadronic molecule.  In the
considered parameter region, the partial widths are evaluated to be
\begin{align}
\Gamma(\Xi_b(6227)\to\gamma\Xi_b)=1.38-1.14~ {\rm KeV},\nonumber\\
\Gamma(\Xi_b(6227)\to\gamma\Xi^{'}_b)=19.47-23.15~ {\rm KeV}.
\end{align}
Our estimations indicated that the partial widths for the $\Xi_b(6227)\to\gamma\Xi_b$ is about one order smaller
than that of $\Xi_b(6227)\to\gamma\Xi^{'}_b$.

Basing on the current integrated luminosity and  our estimations, the facilities like LHCb might have the capability
to detect radiative decays of $\Xi_b(6227)$ baryon in the keV regime.  Such research can also be done in the forthcoming
Belle II experiment.   The study of the radiative decay of $\Xi_b(6227)$ with quark model is strongly recommended.  In
comparison with the predicted widths of quark model, the results in present work can provide further information for the
experimental search for the $\Xi_b(6227)$, and, on the other hand, the experimental measurements for these radiative
decay processes could be a crucial test for the molecule interpretation of the $\Xi_b(6227)$.

\begin{acknowledgments}
This work is partly supported by the Development and Exchange Platform for Theoretic Physics of
Southwest Jiaotong University in 2020(Grants No.11947404).  We acknowledge the supported by the National
Science Foundation of Chongqing (Grant No. cstc2019jcyj-msxm0953), the Science and Technology Research
Program of Chongqing Municipal Education Commission (Grant No. KJQN201800510).
\end{acknowledgments}

\end{document}